\documentclass[aps,prl,nofootinbib,twocolumn,reprint,groupedaddress,showpacs,showkeys]{revtex4-1}

\usepackage{graphicx}  
\usepackage{dcolumn}   
\usepackage{bm}        
\usepackage{amssymb}   
\usepackage{color}

\usepackage {amsmath}
\usepackage {amsfonts}
\usepackage {amsthm} 
\usepackage {mathrsfs}
\usepackage {natbib}
\usepackage {latexsym}
\usepackage {dsfont}
\usepackage {txfonts}
\usepackage {rotating}
\usepackage {wasysym}
\usepackage {multirow}
\usepackage {hhline}
\usepackage {hyperref}
\usepackage {bm}
\usepackage {appendix}
\usepackage {url}
\usepackage {acronym}

\def\commitID{commitID: 45ad6ce16be36be778fd8117904e6e6814f6fd7d}
\def\commitDATE{Thu Dec 1 11:04:23 2011 +0000}

\begin{document}


\title{Measuring a cosmological distance--redshift relationship using only gravitational wave observations of
binary neutron star coalescences}


\author{C.~Messenger}
\affiliation{School of Physics and Astronomy, Cardiff
  University, Queens Buildings, The Parade, Cardiff, CF24 3AA}
\email{chris.messenger@astro.cf.ac.uk}
\author{J.~Read}
\affiliation{Department of Physics and Astronomy, The
  University of Mississippi, P.O. Box 1848, Oxford, Mississippi 38677-1848}


\date{\today}
\date{\commitDATE\\\mbox{\small \commitID}}

\begin{abstract}
  Detection of gravitational waves from the inspiral phase of binary
neutron star coalescence will allow us to measure the effects of the
tidal coupling in such systems. These effects will be measurable using
3$^{\text{rd}}$ generation gravitational wave detectors, e.g. the
Einstein Telescope, which will be capable of detecting inspiralling
binary neutron star systems out to redshift $z\approx 4$.  Tidal
effects provide additional contributions to the phase evolution of the
gravitational wave signal that break a degeneracy between the system's mass
parameters and redshift and thereby allow the simultaneous measurement of both
the effective distance and the redshift for individual sources.
Using the population of $\mathcal{O}(10^{3}\text{--}10^{7})$
detectable binary neutron star systems predicted for the Einstein
Telescope the luminosity distance--redshift relation can be probed
independently of the cosmological distance ladder and independently of
electromagnetic observations. We present the results of a Fisher
information analysis applied to waveforms assuming a subset of
possible neutron star equations of state. We conclude that for our
range of representative neutron star equations of state the redshift of
such systems can be determined to an accuracy of $8\text{--}40\%$ for
$z<1$ and $9\text{--}65\%$ for $1<z<4$.
\end{abstract}


\maketitle

\acrodef{BNS}[BNS]{binary neutron star}
\acrodef{GW}[GW]{gravitational-wave}
\acrodef{EOS}[EOS]{equation of state}
\acrodef{NS}[NS]{neutron star}
\acrodef{EM}[EM]{electromagnetic}
\acrodef{SNR}[SNR]{signal-to-noise-ratio}
\acrodef{ET}[ET]{Einstein Telescope}
\acrodef{GRB}[GRB]{Gamma-Ray burst}
\acrodef{PN}[PN]{post-Newtonian}
\acrodef{LIGO}[LIGO]{Laser Interferometer Gravitational-wave
  Observatory}
\acrodef{ISCO}[ISCO]{innermost-stable-circular orbit}
\acrodef{BHNS}[BHNS]{black-hole---neutron star}
\acrodef{NR}[NR]{numerical relativity}

\emph{Introduction}---
%
%
Making use of \ac{GW} sources as standard sirens (the \ac{GW} analogue
of \ac{EM} standard candles) was first proposed
in~\cite{1986Natur.323..310S}.  It was noted that the amplitude of a
\ac{GW} signal from the coalescence of a compact binary such as a
\ac{BNS} is a function of the redshifted component masses and the
luminosity distance.  Since the former can be estimated separately
from the signal phase evolution, the luminosity distance can be
extracted and such systems can be treated as self-calibrating standard
sirens.  This indicated that \ac{GW} observations do not require the
cosmological distance ladder to measure distances but concluded that
\ac{EM} observations would be needed to measure the redshift of
\ac{GW} sources.  Upon detection of a \ac{GW} signal from a compact
binary coalescence, one could localize the source on the sky using a
network of \ac{GW} detectors.  The host galaxy of the source could
then be identified and used to obtain accurate redshift information
whilst inferring the luminosity distance from the \ac{GW} amplitude.
%
%
%
%
This idea that \ac{GW} and \ac{EM} observations could complement each
other in this way was subsequently extended to include the fact that
\ac{BNS} events are now thought to be the progenitors of most
``short-hard'' \acp{GRB} ~\cite{1989Natur.340..126E}.  The expected
temporal coincidence of these events would allow the more accurately
measured sky position of the \ac{GRB} to be used to identify the host
galaxy.  Recent work~\cite{2010ApJ...725..496N,2010CQGra..27u5006S,
  2011PhRvD..83b3005Z,2011PhRvD..83h4045N} has explored the technical
details regarding the data analysis of \ac{BNS} standard sirens with
respect to the advanced, 3$^{\text{rd}}$ generation ground-based  
\ac{GW} detectors with the aim of investigating the potential of \ac{GW}
observations as tools for performing precision cosmology. The possibility
of cosmological measurements with space-based detectors events is also
promising~\cite{2008PhRvD..77d3512M,2011ApJ...732...82P}.  In addition we
note that statistical arguments based on the assumed \ac{NS} mass
distribution can also be used to infer redshift information from \ac{BNS}
events~\cite{2011arXiv1108.5161T}.  This novel approach is similar to the
work we present here in that it is independent of \ac{EM} counterparts.


The operation of the initial generation of interferometric \ac{GW}
detectors has been successfully completed.  This comprised a network
of four widely-separated Michelson interferometers: the \ac{LIGO}
detectors ~\cite{2010NIMPA.624..223A} in Washington and Louisiana,
USA, GEO600~\cite{2008CQGra..25k4043G} in Hannover, Germany and
Virgo~\cite{2008CQGra..25r4001A} in Cascina, Italy.  We now await the
construction of the advanced detectors~\cite{2010CQGra..27h4006H}
which will recommence operations in ${\sim}2015$ and promise to
provide the first direct detection of \acp{GW}.  It is expected that
in this advanced detector era the most likely first detections will be
from compact binary coalescences of \ac{BNS} systems for which
detector configurations are being tuned~\cite{2005CQGra..22S.461S}.
Astrophysical estimates suggest a rate of detection of at least a few,
and possibly a few dozen, per year~\cite{2010CQGra..27q3001A} with
typical \ac{SNR} $\sim{10}$.  Already much effort has been
spent on the design of a $3^{\text{rd}}$ generation \ac{GW} detector
the \ac{ET}~\cite{2011ETdesigndoc} which is anticipated to be
operational by ${\sim}2025$.  It is designed to be ${\sim}10$ times
more sensitive in \ac{GW} strain than the advanced detectors and as
such we would expect to detect $\mathcal{O}(10^{3}\text{--}10^{7})$
\ac{BNS} events per
year~\cite{2010CQGra..27q3001A,2010CQGra..27u5006S} with \acp{SNR}
ranging up to $\sim{100}$.


In this letter we highlight an important feature associated with the
information that we will be able to extract from \ac{BNS} waveforms
using $3^{\mathrm{rd}}$ generation \ac{GW} interferometers, in
particular \ac{ET}~\cite{2011ETdesigndoc}.  We show that the addition
of the tidal coupling contribution to the \ac{GW} waveform breaks the
degeneracy present in \ac{PN} waveforms between the mass parameters
and the redshift.  This will then allow the measurement of the binary
rest-frame masses, the luminosity distance and redshift simultaneously
for individual \ac{BNS} events.  We base our work on the assumption
that the detections of \ac{BNS} and \ac{BHNS} coalescences made using
both the advanced detectors and \ac{ET} (specifically the nearby high
\ac{SNR} signals) would tightly constrain the universal NS core
\ac{EOS}~\cite{2010PhRvD..81l3016H, 2011arXiv1103.3526P,
  PhysRevD.77.021502, 2009PhRvD..79l4033R}.  Once the \ac{EOS} is
known, the tidal effects are completely determined by the component
rest-frame masses of the system.  Exploitation of these effects would
then remove the requirement for coincident \ac{EM}
observations (so-called ``multi-messenger'' astronomy) to obtain
redshift information. In using \ac{GRB} counterparts for example, host
galaxy identification~\cite{2002AJ....123.1111B} can sometimes be
unreliable,  and we also require that the emission cone from the \ac{GRB}
is coincident with our line of sight.  Current estimates of the
half-opening angles of \acp{GRB} lie in the range
$8\text{--}30^{\circ}$~\cite{2007PhR...442..166N,2011ApJ...732L...6R},
which coupled with the fact that only some short-hard \acp{GRB} have
measured redshifts imply that only a small fraction (${\sim}10^{-3}$)
of \ac{BNS} events will be useful as standard sirens.  Removing the
necessity for coincident \ac{EM} observations will allow all of the
$\mathcal{O}(10^{3}\text{--}10^{7})$ \ac{BNS} events seen with \ac{ET}
to be assigned a redshift measure independent of sky position.  Each
of these detected events provides a measure of the luminosity
distance--redshift relation ranging out to redshift $z\approx 4$.
With so many potential sources the observed distribution of effective
distance (the actual luminosity distance multiplied by a geometric
factor accounting for the orientation of the binary relative to the
detector) within given redshift intervals will allow the accurate
determination of actual luminosity distance and consequently of
cosmological parameters including those governing the dark energy
equation of state.  Such a scenario significantly increases the
potential for 3$^{\text{rd}}$ generation \ac{GW} detectors to perform
precision cosmology with \ac{GW} observations alone.

In our analysis we use a Fisher matrix approach applied to a \ac{PN}
frequency domain waveform to estimate the accuracy to which the
redshift can be measured.  We also assume non-spinning component
masses and treat the waveform as valid up to the \ac{ISCO} frequency,
the implications of which are discussed later in the text.


\emph{The signal model}---We follow the approach
of~\cite{1994PhRvD..49.2658C, 2005PhRvD..71h4008A} in our
determination of the uncertainties in our inspiral waveform
parameters.  We use as our signal model the frequency domain
stationary phase approximation~\cite{1994PhRvD..49.1707D} to the waveform of a non-spinning \ac{BNS}
inspiral,
\begin{equation}\label{eq:phase} 
  \tilde{h}(f) =
  \sqrt{\frac{5}{24}}\pi^{-2/3}Q(\bm{\varphi})\frac{\mathcal{M}^{5/6}}{r} f^{-7/6}e^{-i\Psi(f)},
\end{equation}
where we are using the convention $c=G=1$.  We define the total rest
mass $M=m_{1}+m_{2}$ and the symmetric mass ratio
$\eta=m_{1}m_{2}/M^2$ where $m_{1}$ and $m_{2}$ are the component rest
masses. The chirp mass $\mathcal{M}$ is defined as $\mathcal{M} =
M\eta^{3/5}$, $r$ is the proper distance to the \ac{GW} source and
$\Psi(f)$ is the \ac{GW} phase.  The quantity $Q(\bm{\varphi})$ is a
factor that is determined by the amplitude response of the \ac{GW}
detector and is a function of the nuisance parameters
$\bm{\varphi}=(\theta,\phi,\iota,\psi)$ where $\theta$ and $\phi$ are
the sky position coordinates and $\iota$ and $\psi$ are the orbital
inclination and \ac{GW} polarization angles respectively.  The
standard post-Newtonian point-particle frequency domain phase can be
written as~\cite{2005PhRvD..71h4008A,2005PhRvD..72f9903A}
\begin{equation}\label{eq:PNphase} 
  \Psi_{PP}(f) = 2\pi
  ft_{c}-\phi_{c}-\frac{\pi}{4}+\frac{3}{128\eta
    x^{5/2}}\sum_{k=0}^{N}\alpha_{k}x^{k/2}
\end{equation}
where we use the post-Newtonian dimensionless parameter $x=(\pi
Mf)^{2/3}$ and the corresponding coefficients $\alpha_{k}$ given
in~\cite{2005PhRvD..71h4008A}.  Throughout this work we use $N=7$
corresponding to a 3.5 \ac{PN} phase expansion (the highest known at
the time of publication).  The parameters $t_{c}$ and $\phi_{c}$ are
the time of coalescence and phase at coalescence and we use $f$ to
represent the \ac{GW} frequency in the rest frame of the source.  Note
that if the signal is modeled using the point-particle phase such that
$\Psi(f)=\Psi_{PP}(f)$ then the detected signal $\tilde{h}(f)$ is
invariant under the transformation $(f,\mathcal{M},r,t)\rightarrow
(f/\xi,\mathcal{M}\xi,r\xi,t\xi)$ where $\xi$ is a Doppler-shift
parameter.  For \ac{BNS} systems at cosmological distances the
frequency is redshifted such that $f\rightarrow f/(1+z)$ where $z$ is
the source's cosmological redshift.  Therefore, using the
point-particle approximation to the waveform one is only able to
determine the ``redshifted'' chirp mass
$\mathcal{M}_{z}=(1+z)\mathcal{M}$ and the so-called luminosity
distance $d_{L}=(1+z)r$.  This implies that it is not possible to
disentangle the mass parameters and the redshift from the waveform
alone if the proper distance is unknown.


The leading-order effects of the quadrupole tidal response of a neutron star on
post-Newtonian binary dynamics have been
determined~\cite{2010PhRvD..81l3016H,PhysRevD.83.084051} using 
Newtonian and 1PN approximations to the tidal field. The additional
phase contribution to a \ac{GW} signal from a \ac{BNS} system is given by
\begin{eqnarray}\label{eq:tidalphase} 
  \Psi^{\mathrm{tidal}}(f) &=&\sum_{a=1,2}
\frac{3\lambda_{a}}{128\eta}\left[-\frac{24}{\chi_{a}}\left(1+\frac{11\eta}{\chi_{a}}\right)\frac{x^{5/2}}{M^{5}}\right.
  \\
  &&\left.-\frac{5}{28\chi_{a}}\left(3179-919\chi_{a}-2286\chi_{a}^{2}+260\chi_{a}^{3}\right)\frac{x^{7/2}}{M^{5}}\right]\nonumber
\end{eqnarray}
where we sum over the contributions from each \ac{NS} (indexed by $a$).
The parameter $\lambda = (2/3)R_{\mathrm{ns}}^{5}k_{2}$ characterizes the
strength of the induced quadrupole given an external tidal field, and is a
function of the $l=2$ tidal Love number (apsidal constant) $k_{2}$ for each
\ac{NS}~\cite{2008ApJ...677.1216H,PhysRevD.77.021502}. We  have also
defined $\chi_{a}=m_{a}/M$.  Note that the tidal contributions to the
\ac{GW} phase in Eq.~\ref{eq:tidalphase}  have the frequency dependences of
$x^5$ and $x^6$, and are 5PN and 6PN since when viewed in the context of
the point-particle post-Newtonian phase expansion (Eq.~\ref{eq:PNphase}).
However, for \ac{NS}s, their coefficients are
$\mathcal{O}(R_{\mathrm{ns}}/M)^5{\sim}10^{5}$, making them comparable in
magnitude with the 3PN and 3.5PN phasing terms.

For a chosen universal \ac{NS} \ac{EOS}, 
the perturbation of a spherically symmetric \ac{NS} solution for a given
\ac{NS} mass determines the
NS radius $R_{\mathrm{ns}}$, Love number $k_{2}$ and therefore also
the tidal deformability parameter $\lambda_{a}$.  For the purposes of
this work we use the relationship between the \ac{NS} mass and the tidal
deformability parameter expressed graphically in Fig.~2
of~\cite{2010PhRvD..81l3016H}.  Our approach models this
relationship as a first-order Taylor expansion around the canonical \ac{NS}
mass value such that
$\lambda(m)=\lambda_{1.4}+(d\lambda/dm)_{1.4}(m-1.4M_{\odot})$ where
$\lambda_{1.4}$ and $(d\lambda/dm)_{1.4}$ are the values of the tidal
deformability parameter and its derivative with respect to mass,
both evaluated at $m=1.4M_{\odot}$.

We now highlight the fundamental feature of this work.  With the
addition of the tidal phase components to the total \ac{GW} phase such
that $\Psi(f) = \Psi_{PP}(f)+\Psi_{\mathrm{tidal}}(f)$ the waveform is
no longer invariant under the type of transformation discussed above.
The point-particle \ac{PN} phase as measured at the detector is a
function of the redshifted chirp mass $\mathcal{M}_{z}$ and luminosity
distance $d_{L}$ in contrast to the tidal phase component which
contains terms dependent upon the un-redshifted rest-frame mass
components $m_{1}$ and $m_{2}$.  The degeneracy between the mass
parameters and the redshift is therefore broken and one can now
theoretically measure both sets of quantities independently of one
another.  Essentially, the \ac{NS} size provides a fixed scale-length
that is imprinted on the \ac{GW} waveform.  The ability to perform this measurement is based on the
assumption that one knows or has a very well constrained \ac{NS}
\ac{EOS}.  As shown in~\cite{2009PhRvD..79l4033R}, in the advanced
detector era, departures from the point-particle limit of the \ac{GW}
waveform as the stars approach their final plunge and merger will
place strong constraints on the \ac{EOS}-dependent tidal response of neutron stars.  In the $3^{\text{rd}}$ generation
\ac{GW} detector era, specifically the \ac{ET}~\cite{2011ETdesigndoc},
the subset of high-\ac{SNR} \ac{BNS} signals from local galaxies will
provide even tighter constraints. The addition of future \ac{EM}
observational constraints on the \ac{EOS} (as can be seen currently
in~\cite{2010ApJ...722...33S,2010PhRvD..82j1301O}) will also
contribute to a well-understood \ac{NS} \ac{EOS} by the \ac{ET} era.

The choice of upper cut-off frequency for our model is an important
issue.  The standard approach to tidal effects on \ac{GW} waveforms
has been to use only the Newtonian tidal correction term and to
truncate the signal corresponding to a rest-frame \ac{GW} frequency of
$450$ Hz.  The primary reason that such a choice has been made is to
limit the contributions to the phase evolution from various higher
order effects to ${<}10\%$.  With the addition of the 1PN tidal phase
correction, and neglecting the small known higher-multipole
contributions~\cite{2009PhRvD..80h4035D}, the tidal description is
limited by nonlinear and resonant tidal
effects~\cite{1995MNRAS.275..301K} at the end of
inspiral. Concurrently, the \ac{PN} formalism also breaks down at the
\ac{ISCO} frequency $f_{\text{\ac{ISCO}}} = (6^{3/2}\pi M)^{-1}$
(${\sim}1500$ Hz for $1.4M_{\odot}$ \ac{BNS} systems) where the
secular approximation, that the mode frequency is large compared to
the orbit frequency, also becomes invalid.  
However, recent
\ac{NR}
simulations~\cite{2010PhRvL.105z1101B,2011PhRvD..84b4017B,2011arXiv1109.3611B}
show that \ac{EOS} effects can be accurately modelled in the late inspiral,
and that the waveform contains an \ac{EOS} signature that is amenable to
analytic modelling.
In anticipation of future analytical models valid up to the merger phase, we choose to use the
\ac{ISCO} as our cut-off frequency, noting that this is applied using
un-redshifted mass in the source's local frame.


The standard Fisher matrix
formalism~\cite{1925Fisher,2008PhRvD..77d2001V} allows us to compute
the uncertainties associated with the measurement of a set of signal
parameters. In the large SNR regime under the assumption of Gaussian
noise the signal parameters $\bm{\theta}$ have probability
distribution
$p(\bm{\delta\theta})\propto\exp(-(1/2)\Gamma_{ij}\delta\theta^{i}\delta\theta^{j})$,
where $\delta\theta^{i} = \theta^{i}-\hat{\theta}^{i}$ and
$\hat{\theta}^{i}$ are the best fit parameter values. The Fisher
matrix $\Gamma_{ij}$ is computed via $\Gamma_{ij}=(\partial
h/\partial\theta^{i},\partial h/\partial\theta^{j})$ where the
brackets in this case indicate the noise weighted inner product.  The
expected errors in the measurement of the parameter set $\bm{\delta\theta}$
are then defined by the square root of the diagonal elements of the
inverse Fisher matrix.  We follow~\cite{2005PhRvD..71h4008A} in our
treatment of the parameter estimation analysis for \ac{BNS} \ac{GW}
signals with the addition of the redshift $z$ as a parameter. We
therefore use
$\bm{\theta}=(\ln\mathcal{A},t_{\mathrm{c}},\phi_{\mathrm{c}},\ln\mathcal{M}_{z},\eta,z)$
as our independent parameters where we have absorbed all amplitude
information in to a single parameter $\mathcal{A}$ via
$\tilde{h}(f)=\mathcal{A}f^{-7/6}e^{-i\Psi(f)}$.  The expected SNR of
a given \ac{BNS} signal is dependent upon the nuisance parameters
$\bm{\varphi}$.  For simplicity we have computed our results using the ET-D detector design
configuration~\cite{2011CQGra..28i4013H,2011ETdesigndoc} assuming an
\ac{SNR} that has been
appropriately averaged over each of the 4 constituent angles of
$\bm{\varphi}$.  The detection range of \ac{ET} for \ac{BNS} systems
is $z{\approx}1$
%
%
for such an angle
averaged signal and an \ac{SNR} threshold of 8.  For an optimally oriented
system at the same \ac{SNR} threshold the horizon distance is
$z\approx 4$.  


%
\begin{figure}
  \begin{center}
    \includegraphics[bb = 50 0 320 250,width = \columnwidth]{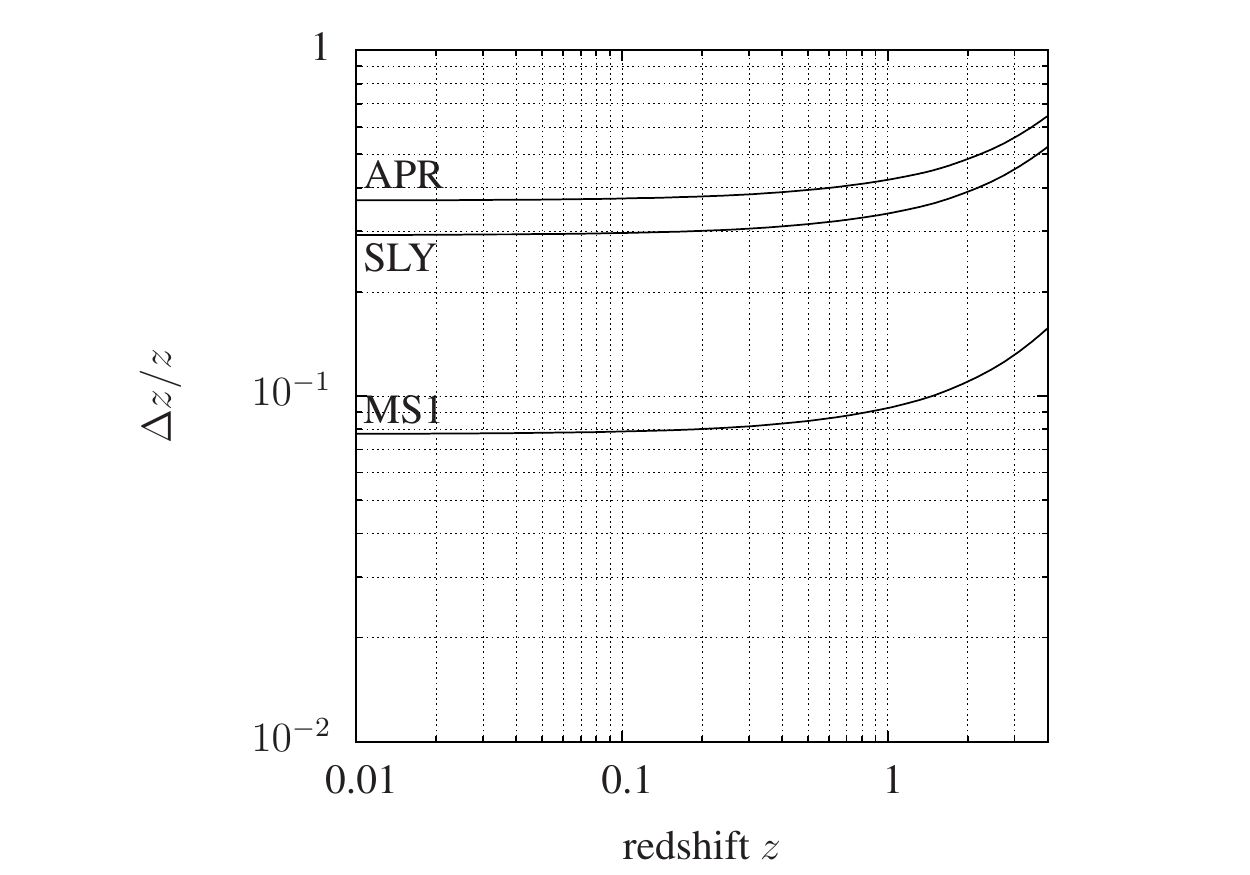}
    \caption{\label{fig:fishererrors} The fractional uncertainties in
      the redshift as a function of redshift obtained from the Fisher
      matrix analysis for \ac{BNS} systems using 3 representative
      \ac{EOS}s, APR~\cite{1998PhRvC..58.1804A},
      SLY~\cite{2001A&A...380..151D} and
      MS1~\cite{1996NuPhA.606..508M}.  In all cases the component
      \ac{NS}s have rest masses of $1.4M_{\odot}$ and waveforms have a
      cut-off frequency equal to the \ac{ISCO} frequency (as defined
      in the \ac{BNS} rest-frame).  We have used a cosmological
      parameter set $H_{0} = 70.5$ kms$^{-1}$Mpc$^{-1}$,
      $\Omega_{m}=0.2736$, $\Omega_{k}=0$,$w_{0} = -1$ to compute the
      luminosity distance for given redshifts and have assumed
      detector noise corresponding to the
      ET-D~\cite{2011CQGra..28i4013H,2011ETdesigndoc} design (a frequency domain
      analytic fit to the noise floor can be found
      in~\cite{ETwebsite}).
      \label{fig:fishererrors_z}}
  \end{center}
\end{figure}
%



\emph{Results}---The results of the analysis with respect to the
uncertainties in the redshift measurement as a function of redshift
are shown in Fig.~\ref{fig:fishererrors} for a subset of \ac{EOS}s.
We have chosen the following 3 (given in order of increasing $\lambda$)
labeled APR~\cite{1998PhRvC..58.1804A}, SLY~\cite{2001A&A...380..151D} and
MS1~\cite{1996NuPhA.606..508M} as representative samples from the set of 18
\acp{EOS} considered in \citet{2010PhRvD..81l3016H}.  The tidal
deformability parameter $\lambda$ for each \ac{EOS} has been parameterized
as a function of the \ac{NS} mass as described above and is directly
proportional to the level of the tidal phase contribution.  
Note that for our most pessimistic choice of \ac{EOS}, the redshift can be
measured to better than $40\%$ accuracy for sources at $z{<}1$ and then
worsens to only $65\%$ at the maximum redshift range, $z{\approx}4$, of ET.
For the most optimistic of our representative\acp{EOS} we see corresponding
values of $8\%$ and $15\%$.  The general trend of the results (for all
\acp{EOS}) is that the fractional redshift uncertainty remains
approximately constant up to $z{\approx}1$ increasing only by a factor of
${\sim}1.5$ for the most distant sources at $z{\approx}4$.  The
relationship between the accuracy of redshift determination and the
\ac{EOS} is, as expected, proportional to the \ac{NS} deformability
parameter $\lambda$.  The consistency of the redshift determination as a
function of redshift can be explained by the combination of 2 competing
effects.  Naively one would expect that the drop in SNR at higher redshifts
would cause any parameter estimation to degrade.  This is true and is the
cause of the final rise in the fractional redshift error at high redshifts.
In parallel, as the more distant sources have their waveforms redshifted to
lower frequencies, the tidal effects which formally begin at 5PN order and
have greatest effect close to the cut-off frequency, are moved towards the
most sensitive band of the detector (${\sim}150$ Hz).  From this argument
one would conclude that this ``sweet spot'' would coincide with $z{\sim}10$
but this effect is diluted at higher redshifts due to a reduction in SNR as
the lower frequency part of the signal moves out of band.



\emph{Discussion}---The analysis presented here is a proof of principle and
is based on a number of assumptions and simplifications which we would
like to briefly discuss and in some cases reiterate. 
%
%
It is likely that by the 3$^{\text{rd}}$ generation \ac{GW} detector
era our knowledge of the tidal response in \ac{BNS} systems will have
significantly advanced through improved \ac{NR}
simulations~\cite{2011PhRvD..83l4008H}.  Current \ac{NR} simulations
have already shown that modelling these tidal phase corrections using
a \ac{PN} formalism, while qualitatively accurate, significantly
\emph{underestimate} the tidal phase
contribution~\cite{2010PhRvL.105z1101B,2011PhRvD..84b4017B,2011arXiv1109.3611B}.
In addition these same studies suggest that it is possible to
accurately model tidal effects up to the merger phase.  Therefore we
feel that our use of the \ac{ISCO} as the upper cut-off frequency of
the \ac{PN} waveforms is a well justified choice for this first estimate.
%
%
We have also neglected the effects of spin in our investigation which
we expect to contribute to the \ac{PN} phase approximation at the level of
${\sim}0.3\%$~\cite{2010PhRvD..81l3016H}. 
This does not preclude the
possibility that marginalizing over uncertainties in spin parameters may 
weaken our ability to determine the redshift. This seems
unlikely given the small expected spins in these systems, as well as the
difference inscalings between the spin terms and the
tidal terms, $x^{-1/2}$ and $x^{5/2}$ respectively, causing the
tidal effects to dominate over spin in the final stage of the
inspiral.  
%
%
We also note that the Fisher information estimate of
parameter uncertainty is valid in the limit of
$\text{SNR}\gtrsim 10$~\cite{2008PhRvD..77d2001V} and under the
assumption of Gaussian noise.  As such, the results at low
SNR, and therefore those at high $z$, should be treated as lower limits via
the Cramer-Rao bound, on the redshift uncertainty. 
%
%
We also mention
here that since the tidal phase corrections are, at leading order,
formally of $5^{\mathrm{th}}$ \ac{PN} order we have uncertainty in the
effect of the missing \ac{PN} expansion terms in the \ac{BNS} waveform between
the 3.5PN and 5PN terms. It is comforting to note that as the \ac{PN} order
is increased our results on the redshift uncertainty do converge to
the point of ${<}1\%$ difference in accuracy between the 3 and
3.5PN terms implying (through extrapolation) that the missing \ac{PN} terms
(as yet not calculated) would not effect our results.  Future detailed
analysis following this work will complement Fisher based estimates
with Monte-Carlo simulations and/or Bayesian posterior based parameter
estimation techniques. 
%
%
Similarly, the signal parameter space should
be more extensively explored beyond the canonical $1.4M_{\odot}$,
equal mass case.  In addition, future work will also include \ac{BHNS}
systems which will also contain, encoded within their waveforms,
extractable redshift information.  Such systems are observable out to
potentially higher redshift although tidal effects will become less important as
the mass ratio increases~\cite{2011arXiv1103.3526P,2011arXiv1109.3402L}.
%
%
Finally, we briefly mention that \ac{GW} detector calibration uncertainties
in strain amplitude (which for $1^{\text{st}}$ generation detectors were
typically ${<}10\%$) will only effect the determination of the luminosity
distance.  Calibration uncertainties in timing typically amount to phase
errors of ${<}1^{\circ}$ and would be negligible in the determination of
the redshift.  Similarly, the effects of weak lensing that would only
affect the luminosity distance measurement have been shown to be negligible
for \ac{ET} sources~\cite{2010CQGra..27u5006S}.


\emph{Conclusions}---Current estimates on the formation rate of \ac{BNS} systems imply that
in the 3$^{\text{rd}}$ generation \ac{GW} detector era there is the
potential for up to ${\sim}10^{7}$ observed events per year out to
redshift $z\approx 4$~\cite{2011ETdesigndoc}.  
%
%
The results presented here suggest that redshift measurements at the
level of ${\sim}10\%$ accuracy can be achieved for \emph{each}
\ac{BNS} event solely from the \ac{GW} observation.  Such systems have
long been known as \ac{GW} standard sirens~\cite{1986Natur.323..310S},
meaning that the luminosity distance can be extracted from the
waveform with accuracy determined by the \ac{SNR} coupled with the ability
with which one is able to infer the geometric orientation of the
source.  Using a large number of sources all sharing the same
redshift, the luminosity distance (free of the orientation parameters)
can be determined statistically from the distribution of observed
amplitudes.  With the ability to extract both the luminosity distance
and the redshift out to such cosmological distances and from so many
sources the precision with which one could then determine the
luminosity distance--redshift relation is significantly enhanced. 
Current proposed methods for making
cosmological inferences using \ac{GW} standard
sirens~\cite{2010ApJ...725..496N, 2011PhRvD..83b3005Z,
2010PhRvD..82f4010M} rely on coincident \ac{EM} counterpart signals
from their progenitors in order to obtain the redshift.  Our method
would allow measurements to be made independently of the cosmological
distance ladder.




\emph{Acknowledgements}---The authors are grateful to J.~Veitch, J.~Clark, R.~Prix, C.~Van~Den
Broeck, B.~S.~Sathyaprakash, P.~Sutton, S.~Fairhurst, M.~Pitkin,
T.~Dent, X.~Siemens, S.~Vitale, L.~Grishchuk and especially J.~Creighton for
 useful discussions and comments. J.~S.~Read is supported by NSF grant PHY-0900735.

\bibliography{../bibtex/masterbib}

\end{document}